\pdfoutput=1
\documentclass[sigconf, natbib, screen=true, anonymous=false, review=false]{acmart}

\acmSubmissionID{arXiv}

\usepackage{amsmath}
\usepackage{amsfonts}
\usepackage{dsfont}
\usepackage{mathtools}
\usepackage{graphicx}
\usepackage{booktabs} 
\usepackage{numprint}
\npthousandsep{,}
\npdecimalsign{.}
\usepackage[inline]{enumitem}
\usepackage{balance}
\usepackage[skip=0pt]{caption}
\usepackage{multirow}
\usepackage{siunitx}
\usepackage{acronym}
\usepackage{caption}
\usepackage{subcaption}
\usepackage{algorithm}
\usepackage[noend]{algpseudocode}
\usepackage{pgfplots}
\usepackage{pgfplotstable} % For handling data
\usepackage{tikz} 
\usetikzlibrary{calc} 
\usetikzlibrary{shapes,arrows.meta,positioning}
\usepackage{{booktabs}}
\usepackage{hyperref}
\PassOptionsToPackage{dvipsnames}{xcolor}

\usepackage{xspace}
\usepackage{soul}

\newcommand{\headernodot}[1]{\vspace{0.8mm}\noindent\textbf{#1}}
\newcommand{\header}[1]{\headernodot{#1.}}

\newcommand\restr[2]{{% we make the whole thing an ordinary symbol
  \left.\kern-\nulldelimiterspace % automatically resize the bar with \right
  #1 % the function
  \vphantom{\big|} % pretend it's a little taller at normal size
  \right|_{#2} % this is the delimiter
  }}

\acrodef{IR}{information retrieval}

\newcommand{\method}{RankingSHAP\xspace}
\newcommand{\greedy}{Greedy\xspace}
\newcommand{\pointwiseShap}{PointwiseSHAP\xspace}

\colorlet{mylimegreen}{yellow!40!green!60!white}

\colorlet{mydarkestgreen}{green!60!blue!80!black}
\colorlet{mydarkgreen}{green!60!blue!90}
\colorlet{mygreen}{green!60!blue!40!white}
\colorlet{myred}{blue!10!red!50}
\colorlet{mydarkred}{blue!20!red!80}
\colorlet{mydarkestred}{blue!30!red!70!black}
\colorlet{mypink}{blue!20!red!30!white}
\colorlet{mydarkpink}{red!50!blue!80}
\colorlet{myorange}{orange!60!yellow!60!white}
\colorlet{mydarkyellow}{green!10!yellow!80!black}
\colorlet{myyellow}{green!10!yellow!80!white}
\colorlet{mydarkorange}{orange!80!yellow!80}
\colorlet{myblue}{red!10!blue!30!white}
\colorlet{mydarkblue}{red!30!blue!70!white}
\colorlet{mybrown}{red!20!brown!70!white}
\colorlet{mydarkbrown}{red!40!brown!90!white}

\allowdisplaybreaks

\copyrightyear{2025}
\acmYear{2025}
\setcopyright{cc}
\setcctype{by}
\acmConference[SIGIR '25]{Proceedings of the 48th International ACM SIGIR Conference on Research and Development in Information Retrieval}{July 13--18, 2025}{Padua, Italy}
\acmBooktitle{Proceedings of the 48th International ACM SIGIR Conference on Research and Development in Information Retrieval (SIGIR '25), July 13--18, 2025, Padua, Italy}\acmDOI{10.1145/3726302.3729971}
\acmISBN{979-8-4007-1592-1/2025/07}

\ccsdesc[500]{Information systems~Evaluation of retrieval results}

\keywords{%
Explainable ranking systems, 
Explainability, 
Explanation evaluation, 
Feature attribution, 
Faithfulness
}

\author{%
Maria Heuss%
}
\orcid{0000-0001-5360-9627}
\affiliation{%
 \institution{
 University of Amsterdam
 \city{Amsterdam}
 \country{The Netherlands}%
 }  
}  
\email{m.c.heuss@uva.nl}

\author{%
Maarten de Rijke%
}
\orcid{0000-0002-1086-0202}
\affiliation{% 
\institution{
 University of Amsterdam
 \city{Amsterdam}
 \country{The Netherlands}%
 }  
}  
\email{m.derijke@uva.nl}

\author{%
Avishek Anand%
}
\orcid{}
\affiliation{%
 \institution{
Delft Institute of Technology
 \city{Delft}
 \country{The Netherlands}%
 }  
}  
\email{Avishek.Anand@tudelft.nl}

\begin{document}

\title[\method{} -- Faithful Listwise Feature Attribution Explanations for Ranking Models]{\method{} -- Faithful Listwise Feature Attribution \\ Explanations for Ranking Models}

\begin{abstract}

While SHAP (SHapley Additive exPlanations) and other feature attribution methods are commonly employed to explain model predictions, their application within information retrieval (IR), particularly for complex outputs such as ranked lists, remains limited.
Existing attribution methods typically provide pointwise explanations, focusing on why a single document received a high-ranking score, rather than considering the relationships between documents in a ranked list. 
We present three key contributions to address this gap.
First, we rigorously define listwise feature attribution for ranking models.
Secondly, we introduce \method, extending the popular SHAP framework to accommodate listwise ranking attribution, addressing a significant methodological gap in the field. 
Third, we propose two novel evaluation paradigms for assessing the faithfulness of attributions in learning-to-rank models, measuring the correctness and completeness of the explanation with respect to different aspects.  
Through experiments on standard learning-to-rank datasets, we demonstrate \method's practical application while identifying the constraints of selection-based explanations. 
We further employ a simulated study with an interpretable model to showcase how listwise ranking attributions can be used to examine model decisions and conduct a qualitative evaluation of explanations.
Due to the contrastive nature of the ranking task, our understanding of ranking model decisions can substantially benefit from feature attribution explanations like \method.
\end{abstract}

\maketitle

\acresetall

\section{Introduction}
\label{ref:intro}

Feature attribution explanations are a posthoc family of explainability approaches that assign scores to features, quantifying their relative contribution to a model's decision. 
They are used to understand which features most influence the model's predictions, thereby enhancing transparency and trust.
Feature attributions are among the most commonly used explanation types for posthoc explanations of trained models in general machine learning (ML)~\cite{ribeiro2016should,kwon2022weightedshap,molnar2023interpreting,zhou2021evaluating}.

Typical ML tasks involve pointwise prediction, explaining single classification or regression decisions. 
However, explaining rankings has different aspects -- Why is a document relevant? (pointwise explanations), Why is one document more relevant than another? (pairwise explanations), or 
Why are the documents ranked in this specific order?
(listwise explanations).
Listwise explanations encode more context in terms of an entire or partial ranked list and are arguably more accurate/faithful since they are able to find features that affect an entire ranking. This is unlike feature attributions that focus on a single relevant document or a certain preference pair.

Feature attribution often lacks rigorous definition, beyond attributing the highest value to the \emph{most important} feature. Limited work exists on pairwise~\cite{penha:2022:pairwise} and listwise explanations~\cite{yu:2022:sigir:generateListExplanation:liege,lyu:2023:ecir:multiplex,singh2020model:fat:pcov,singh2021extracting,anand2023explainable}.
Consequently, listwise feature attribution remains under-explored and in need of further theoretical underpinnings.

\vspace{-0.1cm}
\subsection{A Motivating Case Study -- Talent Search}
To motivate the need for tools that help practitioners arrive at a nuanced understanding of ranking outcomes, we consider talent search. There, systems use learning-to-rank to produce candidate rankings based on features like academic performance, experience, skills, and private attributes such as gender, ethnicity, and university attended. The inclusion of certain attributes in decision-making is debatable, as biases from past decisions can be reflected in the learned model and are best left to human judgment. However, sometimes these attributes are necessary for the model to perform well. Consider the two models in Fig.~\ref{figure:model_flow_chart}. Both use the same features, including skills, experience, graduation grade, university, and whether the candidate meets job requirements. The right model (Fig.~\ref{figure:model_flow_chart-unbiased}) uses the university reasonably by normalizing grades from different institutions, while the left model (Fig.~\ref{figure:model_flow_chart-biased}) discriminates against candidates from certain universities and favors others. Explanations can help differentiate between such models with similar performance to identify which is less biased and more trustworthy. Feature \emph{selection} alone may not provide sufficient insights, as it likely selects the same features ($x_{\text{uni}}$ and $x_{\text{rq}}$) for both models. 
Instead, feature \emph{attribution}, which assigns each feature an importance value, can identify nuanced differences in their relative importance. 
Furthermore, since candidate ranking scores are only meaningful relative to others, \emph{pointwise explanations} focusing on features for high scores may not reveal the university feature as the key factor in determining the relative order for queries with candidates from universities that the model is biased against. 
Pairwise and listwise explanations are better suited to explain relative rankings.
While pairwise explanations require a specification of the pair of candidates to compare, listwise explanations can provide insight into the model decision as a whole. 
We will revisit this case study in Section~\ref{section:experiment-simulated_experiment} to 
demonstrate listwise feature attribution in practice. 

\begin{figure}
\scalebox{0.85}{
    \centering
\subcaptionbox{\label{figure:model_flow_chart-biased}Biased model}
% \resizebox{0.2\textwidth}{!}{
{
\begin{tikzpicture}[
    node distance=0.5cm and 0.1cm,
    mynode/.style={draw,rectangle,align=center},
    mynode1/.style={draw=mydarkorange,rectangle,align=center},
    myarrow/.style={-Stealth},
    edge from parent path={(\tikzparentnode) -- (\tikzchildnode)},
    font=\scriptsize
]
%
% Input feature nodes
\node[mynode] (xskill) {$x_{\text{skill}}$};
\node[mynode, right=of xskill] (xexp) {$x_{\text{exp}}$};
\node[mynode, right=of xexp] (xgrade) {$x_{\text{grade}}$};
\node[mynode, right=of xgrade] (xuni) {$x_{\text{uni}}$};
\node[mynode, right=of xuni] (xrq) {$x_{\text{rq}}$};
%
% Intermediate nodes
\node[mynode, below=of xgrade, xshift=0.2cm, yshift=-0.1cm] (norm) {normalize};
\node[mynode, below=of xexp, yshift=-1cm, xshift=0.3cm] (s) {$s = n + x_{\text{skill}} + x_{\text{exp}}$};
\node[mynode1, below=of s] (cond1) {If \textcolor{mydarkred}{$x_{\text{uni}} = \text{university}_\text{neg-bias}$}};
\node[mynode, below=of cond1, xshift=-0.9cm] (update1) {$s = 0.7 \cdot s$};
\node[mynode1, below=of update1, xshift=0.9cm] (cond2) {If $x_{\text{rq}}$ or \textcolor{mydarkgreen}{$(x_{\text{uni}} = \text{university}_\text{nepotism})$}};
\node[mynode, below=of cond2, xshift=-0.9cm] (update2) {$s = 0.1 \cdot s$};
\node[mynode, below=of update2, xshift=0.9cm](finals) {Final Score $s$};
%
% Paths
\draw[myarrow] (xgrade) -- (norm);
\draw[myarrow] (xuni) -- (norm);
\draw[-Stealth] (xuni) to [bend left=30] (cond1);
\draw[-Stealth] (xuni) to [bend left=20] (cond2);
\draw[-Stealth] (xrq) to [bend left=20] (cond2);
\draw[myarrow] (norm) -- (s);
\draw[myarrow] (xskill) -- (s);
\draw[myarrow] (xexp) -- (s);
\draw[myarrow] (s) -- (cond1);
\draw[myarrow, color=mydarkred] (cond1) -- node[left] {\textbf{Yes}} (update1);
\draw[myarrow] (cond1) -- node[above right] {No} (cond2);
\draw[myarrow] (cond2) -- node[left] {No} (update2);
\draw[myarrow, color=mydarkgreen] (cond2) -- node[above right] {\textbf{Yes}} (finals);
\draw[myarrow] (update1) -- (cond2);
\draw[myarrow] (update2) -- (finals);
\end{tikzpicture}
}
\subcaptionbox{\label{figure:model_flow_chart-unbiased}Unbiased model}
{\begin{tikzpicture}[
    node distance=0.5cm and 0.1cm,
    mynode/.style={draw,rectangle,align=center},
    mynode1/.style={draw=mydarkorange,rectangle,align=center},
    myarrow/.style={-Stealth},
    edge from parent path={(\tikzparentnode) -- (\tikzchildnode)},
    font=\scriptsize
]

% Input feature nodes
\node[mynode] (xskill) {$x_{\text{skill}}$};
\node[mynode, right=of xskill] (xexp) {$x_{\text{exp}}$};
\node[mynode, right=of xexp] (xgrade) {$x_{\text{grade}}$};
\node[mynode, right=of xgrade] (xuni) {$x_{\text{uni}}$};
\node[mynode, right=of xuni] (xrq) {$x_{\text{rq}}$};

% Intermediate nodes
\node[mynode, below=of xgrade, xshift=0.3cm] (norm) {normalize};
\node[mynode, below=of xexp, yshift=-1cm, xshift=0.3cm] (s) {$s = n + x_{\text{skill}} + x_{\text{exp}}$};
\node[mynode1, below=of s] (cond2) {If $x_{\text{rq}}$};
\node[mynode, below=of cond2, xshift=-0.9cm] (update2) {$s = 0.1 \cdot s$};
\node[mynode, below=of update2, xshift=0.9cm](finals) {Final Score $s$};%(update2)!0.5!(update2)$]
\node[mynode, below=of finals, color=white] (white1) {$sDd_{sdf}$};

% Paths
\draw[myarrow] (xgrade) -- (norm);
\draw[myarrow] (xuni) -- (norm);
\draw[myarrow] (norm) -- (s);
\draw[myarrow] (xskill) -- (s);
\draw[myarrow] (xexp) -- (s);
\draw[myarrow] (s) -- (cond2);
\draw[myarrow] (cond2) -- node[left] {No} (update2);
\draw[myarrow] (cond2) -- node[right] {Yes} (finals);
\draw[myarrow] (update2) -- (finals);
\draw[-Stealth] (xrq) to [bend left=25] (cond2);
\end{tikzpicture}}}
    \caption{Flow chart of a biased and an unbiased model for a talent search task. With the help of explanations we would like to be able to differentiate between the two.
    % and pick the unbiased ranker for use in production.
    }
    \label{figure:model_flow_chart}
\end{figure}

\vspace{-0.1cm}
\subsection{Listwise Feature Attribution Explanations}
We are interested in developing a listwise explanation method based on SHAP~\cite{lundberg2017unified}, a method inspired by Shapley values from \textit{game theory}, that quantifies the contribution of each feature to a model's prediction.
SHAP has gained significant popularity as a post-hoc explanation approach due to its theoretical properties and versatility~\cite{krishna2022disagreement}. 
However, SHAP only explains pointwise predictions:
Given the contrastive nature of ranking tasks, listwise feature attribution would provide valuable insights into model decisions by explaining the relative order of documents, enabling comparisons across queries and ranking aspects. To address this gap, we introduce \method, which extends SHAP to support listwise explanations while maintaining compatibility with existing research on SHAP's limitations and extensions. \method provides flexibility in the \emph{listwise explanation objective}, allowing users to determine feature importance for specific ranking aspects that \emph{faithfully} reflect the model's behavior in the context of ranked lists.

\vspace{-0.1cm}
\subsection{Approach and Contributions}
Our proposed method, \method, preserves the context of ranked lists rather than evaluating documents in isolation. This contextual awareness is crucial because ranking models make decisions about relative document ordering. 
Therefore, a feature attribution method needs to identify a specific aspect of the model's decision to focus on and define a singular metric that quantifies changes within the ranked list with respect to that aspect. 
Aspects of interest may include a document's rank, measured by its shift in position, or the overall order of the top-$k$ documents, measured by the number of permutations within the top-$k$.
These diverse aspects underscore the need for a nuanced definition of listwise feature attribution in ranking models, which \method provides.

We rigorously assess the faithfulness of \method using established learning-to-rank (LtR) benchmark datasets, demonstrating its effectiveness in interpreting ranking models' outputs and providing deeper insights into their decision-making processes.

In summary, 
\begin{enumerate*}[label=(\roman*)]
    \item we propose and rigorously define listwise feature attribution;

    \item we present a novel instantiation of our feature attribution framework called \method; and 

    \item we propose multiple evaluation schemes, white box check, preservation and deletion check for ranking feature attributions, and conduct extensive experiments to showcase \method's performance.
\end{enumerate*}

\vspace{-0.1cm}

\section{Related Work}
\label{sec:related-work}

\subsection{Shapley Values and SHAP}
Shapley values, originating from game theory to define a player's marginal contribution~\cite{shapley1953value}, are  widely used in explainable AI. Efficient approximation techniques have facilitated their application in AI model decisions~\cite{strumbelj2010efficient,vstrumbelj2014explaining}. 
SHAP (SHapley Additive exPlanations)~\cite{lundberg2017unified} is one such technique, approximating the expected marginal contribution of a feature to any feature set excluding it. A comprehensive overview and recent advancements are available in~\cite{molnar2023interpreting};
we build on this work, extending it for ranking models. 

Contemporaneously with our work, \citet{pliatsika2024sharp}  propose a Shapley value-based framework for rankings and preferences, but our research emphasizes listwise explanations, unlike their document-level focus. Concurrently, \citet{chowdhury2024rankshap} establish theoretical properties for feature attribution in ranked lists and introduce a method similar to ours that satisfies these properties.

\vspace{-0.1cm}
\subsection{Explainable Information Retrieval}
Explainable IR~\cite{anand2023explainable} has focused on models that are explainable by design~\cite{leonhardt2021learnt,zhang2021explain} and on approaches that can posthoc (after model training) explain models~\cite{Verma:2019:sigir:lirme,Singhexs:2019,singh2020model:fat:pcov}.
Posthoc approaches operate at the global level (model level) or at the local level (per-query). 
Global explainability approaches have been used to diagnose ad-hoc neural text rankers with well-understood axioms of text ranking~\cite{rennings2019axiomatic,camara2020diagnosing:axioms,volske2021towards} or to probe pre-trained transformer-based ranking models for ranking abilities~\cite{wallat:2023:ecir:probing}.
We focus on \emph{posthoc}, \emph{local feature attributions}.

\header{Feature Selection and Attribution for Ranking Models} 
Early work on interpreting ranking models was adapted for explaining query-document relevance from popular paradigms of black-box methods~\cite{ribeiro2016should,lundberg2017unified} or white-box methods~\cite{sundararajan:2017:icml:integratedgradient, simonyan:2013:iclr:saliency, shrikumar:2017:lcml:deeplift}. 
\citet{Singhexs:2019,Verma:2019:sigir:lirme} modify LIME~\cite{ribeiro2016should},
to generate terms as the explanation for a trained black-box ranker. 
\citet{Fernando:2019:sigir:deepshap,choi:2020:arxiv:rankergradcam} applied gradient-based feature attribution methods~\cite{sundararajan:2017:icml:integratedgradient,lundberg2017unified} to interpret document relevance scores.
Contrary to posthoc feature attribution approaches, local feature selection~\cite{lei2016rationalizing,leonhardt2021learnt,hof:sigir:2021:selection:cascading} approaches select a subset of features without distinguishing feature importance.
Most work on local feature selection for rankings~\cite{leonhardt2021learnt,hof:sigir:2021:selection:cascading} is not posthoc, and has been performed on text features, not on learning-to-rank data. 
In this work, we work on posthoc approaches for attribution and not selection.

\header{Listwise Explanations for Ranking Models}
Typical ML tasks are pointwise prediction tasks, i.e., focusing on a single classification or regression decision.
In rankings, even for a single query, we also have to deal with pairwise and listwise explanations, which might be constructed by an aggregation of decisions.
There has been limited work on pairwise~\cite{penha:2022:pairwise} and listwise explanations~\cite{yu:2022:sigir:generateListExplanation:liege,lyu:2023:ecir:multiplex,singh2020model:fat:pcov, saha2024ir_explain}. 
LiEGe~\cite{yu:2022:sigir:generateListExplanation:liege} tackles the task as text generation.  
Other work uses simple rankers to approximate the original ranking of a complex black-box model by expanding query terms by solving a combinatorial optimization problem~\cite{singh2020model:fat:pcov,lyu:2023:ecir:multiplex}.
The work that is closest work to ours, on RankLIME~\cite{chowdhury2023rank}, approaches the problem with the local surrogate approach LIME, which the authors adapt for ranking models. 
Again, most of the approaches focus on text features and are not directly applicable to learning-to-rank models.

\header{Explainability in Learning-to-Rank}
Local feature selection approaches can be applied to learning-to-rank~\cite{GAS-geng2007feature,gigli2016fast,purpura2021neural}.
Among the feature-selection approaches, filter methods are model-agnostic~\cite{GAS-geng2007feature}, while wrapper methods are designed for a particular type of model \cite{gigli2016fast}.
In the context of ranking, some work produces local feature selections~\cite{singh:2018:arxiv:secondaryData,purpura2021neural}.
\citet{singh:2020:ictir:validLTR} proposes the notions of validity and completeness based on the information contained in the explanation.
While these notions are useful in both conception and evaluation of explanations, they still view the explanation as a \emph{selection} of features.
Feature selection methods, however, lack the capability to differentiate between features of varying importance, thereby avoiding a nuanced understanding of which features are substantially more critical in the decision-making process. We focus on feature attributions.

\vspace{-0.1cm}
\subsection{Faithfulness in Explainable AI}
\label{sec:faithfulnes}

Faithfulness measures how accurately an explanation represents the reasoning process behind a model's prediction~\cite{jacovi2020towards}. 
Evaluating faithfulness is challenging because the model's actual reasoning cannot be directly observed. 
Hence, various definitions and evaluation frameworks for faithfulness have been proposed~\cite{jacovi2020towards,lyu2024towards}.
While there is no clear agreement as to what notion or framework should be used to measure and establish faithfulness~\cite{lyu2024towards}, there are two dominant frameworks in explainable IR~\cite{anand2023explainable}.
When locally approximating a ranking model with a proxy model, faithfulness is the degree to which the proxy model approximates the original ranking~\cite{singh2020model:fat:pcov,lyu:2023:ecir:multiplex}. An alternative notion of faithfulness is based on an \textit{information-theoretic} notion of feature importance~\cite{wiegreffe-pinter-2019-attention,singh:2020:ictir:validLTR}.
There, faithfulness refers to the predictive power of the features in the attribution.
Specifically, if a feature set is important then masking off or removing the non-relevant features should not result in a big change in model output.
While both notions model different aspects of faithfulness, in this paper we follow the latter framework.

%!TEX root = ./main.tex

\section{Feature Attribution for Pointwise Rankers}

Early work on local feature explanations has introduced the concept of feature attribution~\cite{strumbelj2010efficient}; recent work often lacks a clear definition of what makes a feature \emph{important}, causing ambiguity in evaluating attribution faithfulness. Despite attempts to formalize feature attribution \cite{afchar2021towards}, these efforts have not been widely adopted, resulting in inconsistencies and confusion in the field~\cite{krishna2022disagreement}.
We build on~\cite{lundberg2017unified} to define  \emph{pointwise feature attribution} for black-box models with one-dimensional model output such as a pointwise ranking model
\begin{equation}
\tilde R: \mathcal{D} \to \mathbb{R}, x_{q,l}\mapsto s_{q,l},
%\text{, i.e. } s_{q,l}:= R(x_{q,l}).
\end{equation}
that predicts the ranking scores $s_{q,l}\in \mathbb{R}$, representing the probability of relevance, for the feature vectors of each document-query pair, $x_{q,l} \in \mathcal{D}$ in the space of all documents $\mathcal{D}$. We consider \emph{instance-wise} feature attribution explanations that assign to each feature $i$ an attribution value $\phi_i(x, \tilde R)$, directly reflecting the importance of the feature to the model decision for instance $x$. Hence, \emph{feature attribution explanations} can be understood as dictionaries $\{i \mapsto \phi_i(x, \tilde R)\}_{i=1,\dots , n}$ containing exactly one attribution value per feature.
A well-defined, instance-specific definition of feature attributions should consider the specific combinations of feature values in the input that collectively lead the model to predict a high score. Also, features with greater importance for the prediction should have higher attribution values.

\label{section:definition_definition}
We use marginal contributions to define \emph{pointwise feature attribution}.\footnote{For a detailed discussion of marginal contributions, see \citep{molnar2023interpreting}.} Our definition is based on SHAP~\citep{lundberg2017unified}. In Section~\ref{section:method}, we extend this to \emph{listwise feature attribution} and define \method to approximate feature attribution for listwise rankers.

\begin{definition}
We define the \emph{attribution} or \emph{importance} of a feature $j$ in terms of marginal contributions. Let $n=dim(\mathcal{D})$ be the input space dimension, and let a \emph{coalition} be a subset $S\subset \{1,\dots, n\} \setminus j$ of the input features excluding $j$. To measure the marginal contribution of feature $j$ to coalition $S$, we compare the model output when shown only features in $S$ to the output when shown features in $S\cup \{j\}$. Since we cannot simply erase features, we mask them with samples from a set of feature-vectors $B\subset \mathcal{D}$, called \emph{background data}, which ideally summarizes the data distribution. For masking, we use templates defined by subsets $S$, indicating the presence $(i\in S)$ or absence $(i\notin S)$ of a feature, and data-points from the background data $b\in \mathcal{D}$. We define $m_{S,b}:\mathcal{D} \to \mathcal{D}$ as:
\begin{align}
\label{Equation:masking}
m_{S,b}(x)_i = 
\begin{cases}
    x_i ,& \text{if } i\in S\\
    b_i, & \text{if } i \notin S.
\end{cases}
\end{align}
The marginal contribution of feature $j$ to coalition $S$ for %background 
vector $b$ is:
\begin{equation}
    \tilde R(m_{S\cup\{j\}, b}(x)) - \tilde R(m_{S, b}(x)).
\end{equation}
We define the \textbf{pointwise feature attribution} of feature $j$ to the model decision of $\tilde R$ at input $x$ as the expected marginal contribution of feature $j$ to all possible coalitions of features:
\begin{align*}
   \phi_j(x,\tilde R) =  \sum_{S\subset \{1,\dots n\}\setminus j} w_S \cdot \mathbb{E}_{b\sim B} [ \tilde R(m_{S\cup\{j\}, b}(x)) - \tilde R(m_{S, b}(x))],
    \label{formula:marginal_contribution}
\end{align*}
with weighting factor $w_S =  \frac{1}{n!}{|S|! (n-|S|-1)!} $  and uniform sampling from $B$. 
\end{definition}

\header{Computational Costs} Given the exponential growth of coalitions with the number of features and the need for numerous background examples for a good summary, we approximate pointwise feature attribution using sampling. Following \cite{lundberg2017unified}, we use SHAP for this approximation. Even though we are approximating the attribution values, SHAP is known to be computationally expensive, especially for high feature dimensions. There have been advances to making the sampling more efficient~\cite{jethani2021fastshap, zhang2023efficient}. Also, since pointwise explanations are usually used as an analysis tool for specific input examples rather than to analyze the whole corpus, it remains a broadly used explanation approach~\cite{molnar2023interpreting, krishna2022disagreement} despite its computational costs.

\vspace*{-2mm}
\section{Feature Attribution for Listwise Rankers}\label{section:method}

For many machine learning tasks, SHapley Additive exPlanations (SHAP)~\cite{lundberg2017unified} effectively approximate feature attribution values for individual model decisions, such as regression scores or classification probabilities. However, applying this method to listwise ranking models is challenging because these models output a ranked list rather than a single score. 
Within this ranked list, different decisions are made regarding the order of individual documents. Pointwise SHAP is only defined for a single one-dimensional model output. While it can explain the model score of an individual document, it does not consider the context of other documents in the list. 
In this work, we extend SHAP to an approach that caters to listwise ranking decisions, called \method.

Instead of looking at pointwise ranking models, as we did in Section~\ref{section:definition_definition}, we consider a listwise ranking model 
\begin{equation}
    R: \{\mathcal{D}_q\}_q \to \text{Sym}, \{x_{q,j}\}_j \mapsto \pi_{q}
\end{equation}
that maps a set of candidate feature vectors for query $q$, $\mathcal{D}_q=\{x_{q,j}\}_{j}$, to some permutation matrix $\pi_{q}\in \text{Sym}(\mathcal{D}_q)$ representing the ranked list in the Symmetry group of all permutations of the candidate set $\mathcal{D}_q$. 

We define two components, \emph{listwise masking} and \emph{listwise explanation objectives} that enable us to establish listwise feature attribution for ranking models, which we will introduce in Section~\ref{section:method_idea}. In Section \ref{section:method_formal_definition}, we formally define \method for approximating listwise attribution values. We define \method as a wrapper around SHAP using those two components. We deliberately chose not to modify SHAP's internal algorithm, allowing us to leverage the extensive literature on SHAP directly. Finally, we examine listwise explanation objectives with examples in Section~\ref{section:method_explanation_objectives}.

\subsection{Feature Attribution for Ranking Models}
\label{section:method_idea}
Our definition of feature attribution/feature importance for ranking models consists of two parts: 
\begin{enumerate*}[label=(\roman*)]
\item Define how masking applies to each document in the ranking $\mathcal{D}_{q}$ for query $q$. And
\item measure the impact of input changes on the model decision, quantified by a single number.
\end{enumerate*}

\header{Masking the Inputs of a Ranking Model}
We apply a listwise mask $m_{S,b}$ to all documents $\{x_{q,j}\}_j$ in the ranking:
$m_{S,b}(\mathcal{D}_q) = \prod_{x_{q,j} \in \mathcal{D}_q}m_{S,b}(x_{q,j})$.
By masking the feature vector $x_{q,j}$ of each document with the same mask $m_{S,b}$, we disregard the impact of the masked features on the ranking decision. This helps identify the contributions of non-masked features to the document ordering.

\header{Reducing the Model Prediction to a Single Prediction Value}
Feature attribution is defined by the expected change in the predicted score. We need to reduce the ranking model's decisions to a single value reflecting the change for a perturbed input sample, using a listwise explanation objective that takes a ranked list and maps it to a value, highlighting some property of the ranked list that we want to investigate. 

One example for such a function is a rank similarity coefficient like Kendall's tau $\tau$~\cite{kendall-1938-new}, which is commonly used in the interpretability literature to measure rank correlation~\cite{lyu:2023:ecir:multiplex,singh:2020:ictir:validLTR,singh2020model:fat:pcov}. 
By comparing the change in the relative order of documents, we can measure how much the prediction deviates from the optimal order $\pi_q$ predicted by the model:
\begin{equation}\label{equation:objective_kendall}
g_q(\tilde \pi) = \tau ( \pi_q, \tilde \pi),
\end{equation}
For any such \emph{listwise explanation objective} $g_q$, we define feature importance through the composition with the original ranking model, $g_q \circ R$.  Section~\ref{section:method_explanation_objectives} provides further examples. 

In summary, we have defined how to ``remove'' a feature from the model input through masking and measure its impact on the model prediction with a single value. This allows us to determine the \textbf{listwise feature attribution} using Section~\ref{section:definition_definition}.

\subsection{Estimating Listwise Feature Attribution with \method}\label{section:method_formal_definition}

With the definition of feature attribution for ranking models, we introduce \method. This depends on the choice of listwise explanation objective $g$ and aims to explain which features are important for specific aspects of the ranked list. The ability to focus on different aspects of the ranking decision allows \method to provide contrastive and flexible instance-wise explanations for rankers.

Following the definition of feature attribution with simultaneous masking of document vectors and a listwise explanation objective, we establish \method as a wrapper around SHAP to approximate the marginal contribution of each feature in a ranking model, leveraging prior work.

SHAP samples both coalitions (templates for creating masks) and background data to generate masked perturbations (see Eq.~\ref{Equation:masking}) of the input, approximating the marginal contribution of a feature to any coalition. Given a sampled mask $m_{S,b}$, we illustrate how \method adjusts the model prediction for use with SHAP in Algorithm~\ref{alg:rankingshap}. We loop over all documents $x_j \in \mathcal{D}_q$ (lines \ref{alg:ranking_shap_line_1}--\ref{alg:ranking_shap_line_3}) and perturb the document features with the mask to get $\tilde x_j = m_{S,b}(x_j)$. Then, we rank the perturbed feature vectors with the ranking model $\pi = R(\{\tilde x_j\}_j)$ (line \ref{alg:ranking_shap_line_4}). Finally, we apply the listwise explanation objective $v = g(\pi)$ to measure the change in output according to the specified explanation objective (lines \ref{alg:ranking_shap_line_5} and \ref{alg:ranking_shap_line_6}).

\header{Computational Costs} Our approach allows for the use of existing SHAP implementations. This also means that it inherits any limitation that SHAP has such as the computational complexity. Nevertheless, it does not introduce any significant new additional computational overhead and allows us to use prior research on SHAP extensions and improvements for ranking without adjustments, such as advances in improving efficiency. Since SHAP is a commonly used explanation approach for pointwise predictions, we do not expect the computational complexity of \method to hinder it's adoption in practice.

\algrenewcommand\algorithmicensure{\textbf{Input:}}
\begin{algorithm}[t] 
% \small
\caption{Adjusted model prediction (used in combination with SHAP)}
\label{alg:rankingshap}
\begin{algorithmic}[1]
\Require  ranking-model $R$, feature-vectors $\mathcal{D}_q$ for query $q$, listwise explanation objective $g$, 
\Ensure masking function $m_{S,b}$
\ForAll{$x_j\in \mathcal{D}_q$} \label{alg:ranking_shap_line_1}
    \State $\tilde x_j \gets m_{S,b}(x_j)$\label{alg:ranking_shap_line_2}
\EndFor\label{alg:ranking_shap_line_3}
\State $\pi \gets R(\{\tilde x_j\}_j)$\label{alg:ranking_shap_line_4}
\State $v \gets g(\pi)$ \label{alg:ranking_shap_line_5}
\State \textbf{return} $v$\label{alg:ranking_shap_line_6}
\end{algorithmic}
\end{algorithm}

\subsection{Listwise Explanation Objectives}\label{section:method_explanation_objectives}
We provide examples of listwise explanation objectives to illustrate the types of contrastive explanations \method can generate.

\header{Emphasizing Top-Ranked Documents}
Instead of focusing on the entire ranked list, we can emphasize the top-$k$ documents to identify features crucial for their high ranking. For example, we demonstrate \method using a weighted rank difference objective with common position weighting:
\begin{equation}\label{equation:objective_weighted}
g^w_{q}(\tilde \pi ) = \sum_{d\in D_q}{\frac{\operatorname{rank}(d |\tilde \pi) - \operatorname{rank}(d | \pi_q) }{\operatorname{log}_2(\operatorname{rank}(d | \pi_q))}}.
\end{equation}

\header{Explaining Feature Importance of a Singular Document}
This objective focuses on one particular document $d$, investigating which features contribute, or would contribute, most to its high ranking compared to others when only a subset of features is considered. This can be implemented using the negative rank\footnote{We use the negative rank to maintain consistency with higher values being more desirable, explaining why a document ranks high (low rank) rather than low.} of that document:
\begin{equation}\label{equation:objective_one_vs_all_rank}
g^{\operatorname{rank}(d)}_{q}(\tilde \pi ) = - \operatorname{rank}(d |\tilde \pi).
\end{equation}
Alternatively, we can use \method to determine the features that are the most beneficial for the document's exposure: 
\begin{equation}\label{equation:objective_exposure}
g^{\operatorname{exp}(d)}_{q}(\tilde \pi ) = \operatorname{exp}(\operatorname{rank}(d |\tilde \pi)) = 1 / \operatorname{log}_2(\operatorname{rank}(d | \tilde\pi)).
\end{equation}

\header{Explaining the Position of a Group of Documents}
\method allows us to compare ranking decisions for two groups of documents. We can consider the relative ordering or absolute distance of members of the different groups. Future work could explore explaining model fairness or identifying biases using listwise feature attribution.
\section{Talent Search: A White Box Example}
\label{section:experiment-simulated_experiment}

To demonstrate the application of \method and to evaluate the feature attributes generated by different explanation approaches, we create a synthetic example, revisiting the talent search case study from the introduction. We design an interpretable model to estimate the importance of features for various model decisions. This evaluation framework, known as a ``White Box Check,'' is widely used in the explainability community for other ML tasks~\cite{nauta2023anecdotal}.

In the following sections, we define features and ranking model that we will use as white box in Section~\ref{section:experiments-model}. We then describe the experimental setup in Section~\ref{section:experiments_simulated_evaluation} and examine various queries modeling different types of model decisions in Section~\ref{section:experiments_scenarios}. These queries demonstrate the practical use of listwise feature attribution and qualitatively evaluate three feature explanation approaches. In Section~\ref{section:toy-example-zoom-in}, we show how to use \method to zoom in on individual documents and compare it to a pointwise explainer. We conclude with a detailed discussion in Section~\ref{section:experiment-simulated-discussion}.

\subsection{Model Design}\label{section:experiments-model} 
We design a model using 5 features indicating whether a candidate meets general job requirements, the university the candidate graduated from, skill and experience levels, and average graduation grade. This model ranks candidates for various academic degree-required scenarios, aiming to mimic biases in trained models.

\begin{table}[!ht]
\caption{Candidate evaluation criteria for running example}
\label{table:simulated_experiment_features}
%\footnotesize
% \small
\centering
\begin{tabular}{>{\bfseries}p{0.2\columnwidth}p{0.7\columnwidth}}
\toprule
Feature & \textbf{Description} \\
\midrule
Requirements & Binary value $x_{\text{rq}}\in\{\text{T}, \text{F}\}$ indicating if the candidate meets the job's minimum requirements. \\
Experience & Relevant work experience on a scale $x_{\text{exp}}\in [0,1]$ (1=extensive experience, 0=none) \\
Skills & Skill fit on a scale $x_{\text{skill}}\in [0,1]$, (1 = perfect match, 0 = no relevant skills) \\
University & Institution where the candidate obtained their degree, $x_{\text{uni}}$. \\
Grades & Mean graduation grade, $x_{\text{grade}}$, with range depending on the university. \\
\bottomrule
\end{tabular}
\end{table}
Detailed feature information is in Table~\ref{table:simulated_experiment_features}.
The model favors candidates from $\textbf{uni}_\textbf{nepotism}$ and disadvantages those from $\textbf{uni}_\textbf{neg-bias}$. A flowchart is in Fig.~\ref{figure:model_flow_chart-biased} in Section~\ref{ref:intro}. The ranking score is determined as follows:

\begin{itemize}[leftmargin=*,nosep]
    \item Normalize the grade $\textbf{norm}(x_{\text{grade}}, x_{\text{uni}})$, scaling it so that the minimum possible grade is 0 and the maximum is 1, to make grades from universities with different grading schemes comparable.
    \item Calculate the sum of $x_{\text{skill}}$, $x_{\text{exp}}$, and $\text{norm}(x_{\text{grade}}, x_{\text{uni}})$.
    \item For candidates from $\textbf{university}_\textbf{neg-bias}$, apply a negative bias by multiplying the score by 0.9.
    \item If the candidate does not meet the job requirements, multiply the score by 0.25, effectively placing them at the bottom of the list. Candidates from $\textbf{university}_\textbf{nepotism}$ are exempt from this penalty.
\end{itemize}
\noindent%
Candidates are ranked by their scores, with the highest at the top. We then investigate different queries with \method to identify biases and compare attribution values to other explanation approaches.

\vspace{-1mm}
\subsection{Experimental Setup}
\label{section:experiments_simulated_evaluation}

The main goal of this Section is to showcase the usage of \method and demonstrate the need for listwise, as opposed to pointwise, explanations and feature attribution rather than feature selection. Therefore, we compare \method to the pointwise SHAP explainer, \textbf{\pointwiseShap} (averaged over all candidates), as well as to the \textbf{\greedy} feature selection approach from \cite{singh2021extracting}. The latter iteratively adds features to an initially empty set based on their marginal contribution to the Kendall's tau objective from Eq.~\ref{equation:objective_kendall} until the contribution becomes non-positive or the explanation size reaches 2.
Section~\ref{section:experiments} contains a more complete empirical comparison with a comprehensive set of baselines, including RankLIME~\cite{chowdhury2023rank} and ShaRP~\cite{pliatsika2024sharp}. For background data, we sample 100 candidates from uniform distributions over the possible feature values defined in Section~\ref{section:experiments-model}. Detailed feature values and candidate lists for each query are provided in Appendix~\ref{app:section:experimental-setup}.

\vspace{-1mm}
\subsection{Listwise Evaluation Across Query Scenarios}
\label{section:experiments_scenarios}

We define scenarios to demonstrate feature attribution for contrastive ranking explanations and evaluate them. We present 5 query scenarios: three in the main body and two in Appendix B%~\ref{section:appendix_additional_query_scenarios}
.\footnote{An extended appendix including these additional results is available at \url{https://github.com/MariaHeuss/RankingShap/blob/main/Paper_RankingSHAP.pdf}.} %\footref{footnote_appendix} 
We discuss the setup, candidate constellation, estimated feature importance $imp_{\textit{feature}}$ for some features on the overall ranking, and evaluate the explanation approaches. In this part of our analysis \method uses Kendall's tau explanation objective from Eq.~\ref{equation:objective_kendall} to explain the overall order of the candidates.

\vspace{-1mm}
\subsubsection{Average query}
\header{Description} This query includes candidates from universities with the same grading scheme, only some meeting the requirements, none from $\text{university}_\text{neg-bias}$ or $\text{university}_\text{nepotism}$.

\header{Importance}
Since no exceptions for candidates from biased institutes apply and grades are within the same scheme, we expect $imp_{\text{rq}}$ to be high, as hiding this feature could change the ranking significantly. We also expect $imp_{\text{uni}}$ to have a positive but smaller value since a change of university for all candidates causes ambiguity for the evaluation of the grade. 

\header{Evaluation of Feature Attributes}
Fig.~\ref{figure:synthetic_bar_in_paper}(a) shows attribution values/selected features (bars with length 1). Both \method and \pointwiseShap identify $x_{\text{rq}}$ as an important feature and assign a positive value to $x_{\text{uni}}$. The greedy feature selection approach only selects the university feature.

\begin{figure}
\centering
% \small

\newcommand{\SpacingX}{0.2em}
\newcommand{\Width}{0.2\textwidth}
\newcommand{\Height}{0.21\textwidth}
\newcommand{\BarWidth}{0.0035\textwidth}
\newcommand{\BarWidthI}{0.009\textwidth}
\newcommand{\BarShift}{0.0055\textwidth}

\begin{tikzpicture}
\begin{axis}[
    xbar,
    width=\Width, 
    height=\Height,
    xlabel={},
    symbolic y coords={experience,skills,grades,university,requirements},
    ytick={experience,skills,grades,university,requirements},
    yticklabels={exp,skills,grad,uni,req},
    nodes near coords align={horizontal},
    xtick={-0.5,0,0.5}, 
    xmin=-1, 
    xmax=1,
    ytick style={draw=none}, 
]
\addplot [
    draw=mydarkgreen, % Border color for bars
    fill=mygreen,    
    bar width=\BarWidth, % Adjust the bar thickness here
    bar shift=-\BarShift, % Shift the first set of bars up
] table [
    col sep=comma,
    x=average,
    y=features,
] {figures/csv_files/rankingshap_no_weighting_syn_u.csv};
\addplot [
    draw=mydarkorange, % Border color for bars
    fill=myorange, % Default fill color for bars
    bar width=\BarWidth, % Adjust the bar thickness here
    % bar shift= - (2 * \BarShift + \BarWidth, % Shift the first set of bars up
] table [
    col sep=comma,
    x=average,
    y=features,
] {figures/csv_files/aggregated_shap_top_5_syn_u.csv};
\addplot [
    draw=mydarkblue, % Border color for bars
    fill=myblue, % Default fill color for bars
    bar width=\BarWidth, % Adjust the bar thickness here
    bar shift=\BarShift
] table [
    col sep=comma,
    x=average,
    y=features,
] {figures/csv_files/greedy_listwise_iter_syn_u.csv};
% \addplot [
%     draw=mydarkred, % Border color for bars
%     fill=myred, % Default fill color for bars
%     bar width=\BarWidthI, % Adjust the bar thickness here
% ] table [
%     col sep=comma,
%     x=average,
%     y=features,
% ] {figures/csv_files/importance_nan.csv};
\end{axis}
\node[above,font=\large\bfseries, xshift=6] at (current bounding box.north) {(a) Average};
\node[below, color=white] at (current bounding box.south) {Attribution values};
\end{tikzpicture}\hspace{-0.4cm} 
\begin{tikzpicture}
\begin{axis}[
    xbar,
    width=\Width, height=\Height,
    xlabel={},
    symbolic y coords={experience,skills,grades,university,requirements},
    ytick=data,
    nodes near coords align={horizontal},
    xtick={-0.5,0,0.5}, 
    xmin=-1, 
    xmax=1,
    yticklabels={},
    ytick={},
    ytick style={draw=none}, 
    % nodes near coords={\pgfmathprintnumber{\rawvalue}}, % Print the value near coords
]
\addplot [
    draw=mydarkgreen, % Border color for bars
    fill=mygreen,    
    bar width=\BarWidth, % Adjust the bar thickness here
    bar shift=-\BarShift, % Shift the first set of bars up
] table [
    col sep=comma,
    x=qualified,
    y=features,
] {figures/csv_files/rankingshap_no_weighting_syn_u.csv};
\addplot [
    draw=mydarkorange, % Border color for bars
    fill=myorange, % Default fill color for bars
    bar width=\BarWidth, % Adjust the bar thickness here
] table [
    col sep=comma,
    x=qualified,
    y=features,
] {figures/csv_files/aggregated_shap_top_5_syn_u.csv};
\addplot [
    draw=mydarkblue, % Border color for bars
    fill=myblue, % Default fill color for bars
    bar width=\BarWidth, % Adjust the bar thickness here
    bar shift=\BarShift,
] table [
    col sep=comma,
    x=qualified,
    y=features,
] {figures/csv_files/greedy_listwise_iter_syn_u.csv};
% \addplot [
%     draw=mydarkred, % Border color for bars
%     fill=myred, % Default fill color for bars
%     bar width=\BarWidthI, % Adjust the bar thickness here
% ] table [
%     col sep=comma,
%     x=qualified,
%     y=features,
% ] {figures/csv_files/importance_nan.csv};
\end{axis}
\node[above,font=\large\bfseries] at (current bounding box.north) {(b) Qualified};
\node[below,xshift=5] at (current bounding box.south) {Attribution values};
% \node[below, color=white] at (current bounding box.south) {Attribution values};
\end{tikzpicture}\hspace{-0.55cm}%
\begin{tikzpicture}
\begin{axis}[
    reverse legend,
    xbar,
    width=\Width, height=\Height,
    xlabel={},
    symbolic y coords={experience,skills,grades,university,requirements},
    ytick=data,
    nodes near coords align={horizontal},
    xtick={-0.5,0,0.5}, 
    xmin=-1, 
    xmax=1,
    yticklabels={},
    ytick={},
    ytick style={draw=none}, 
    legend style={at={(1.55,0.3)},anchor=east, font=\small},  
    legend image code/.code={
            \draw [#1] (0cm,-0.1cm) rectangle (\BarWidthI, \BarWidth); },
    ]
]
\addplot [
    draw=mydarkgreen, % Border color for bars
    fill=mygreen,    
    bar width=\BarWidth, % Adjust the bar thickness here
    bar shift=-\BarShift, % Shift the first set of bars up
] table [
    col sep=comma,
    x=biased,
    y=features,
] {figures/csv_files/rankingshap_no_weighting_syn_u.csv};\label{rankshap}
\addplot [
    draw=mydarkorange, % Border color for bars
    fill=myorange, % Default fill color for bars
    bar width=\BarWidth, % Adjust the bar thickness here
] table [
    col sep=comma,
    x=biased,
    y=features,
] {figures/csv_files/aggregated_shap_top_5_syn_u.csv};\label{aggrshap}
\addplot [
    draw=mydarkblue, % Border color for bars
    fill=myblue, % Default fill color for bars
    bar width=\BarWidth, % Adjust the bar thickness here
    bar shift=\BarShift,
] table [
    col sep=comma,
    x=biased,
    y=features,
] {figures/csv_files/greedy_listwise_iter_syn_u.csv};\label{greedy}
% \addplot [
%     draw=mydarkred, % Border color for bars
%     fill=myred, % Default fill color for bars
%     bar width=\BarWidthI, % Adjust the bar thickness here
% ] table [
%     col sep=comma,
%     x=biased,
%     y=features,
% ] {figures/csv_files/importance_nan.csv};\label{importance}
\legend{RShap, PWShap, Greedy}
\end{axis}
\node[above,font=\large\bfseries,xshift=-0.3cm] at (current bounding box.north) {(c) Neg biased};
\node[below, color=white] at (current bounding box.south) {Attribution values};
\end{tikzpicture}

\caption{Feature attribution values for different query scenarios from Section \ref{section:experiments_scenarios}\label{figure:synthetic_bar_in_paper}.}
\label{fig:RQ3}
\end{figure}

\vspace{-1mm}
\subsubsection{Qualified query} 
\header{Description} Similar to the average query, but only candidates meeting the requirements. The model can ignore $x_{\text{rq}}$ without bias.

\header{Importance} 
While $imp_{\text{uni}}$ should still be assinged a positive value, $imp_{\text{rq}}$ should be assigned a lower value than before 
as $x_{\text{rq}}$ is irrelevant for these candidates.

\header{Evaluation of the feature attributes}
Fig.~\ref{figure:synthetic_bar_in_paper}(b) shows that \greedy and \method correctly assign a low value to the $x_{\text{rq}}$. 
\pointwiseShap is not able to identify that the feature that is most important for attaining a high ranking score for each individual document, $x_{\text{rq}}$, is not important for this specific query. Furthermore, we notice that \method assigns higher values to other features, that are now important to distinguish between the candidates. 

\vspace{-1mm}
\subsubsection{Negative bias query} \label{section:query_scenario_bias}
\header{Description} 
Similar to the average query, with an additional candidate from $\text{university}_\text{neg-bias}$ having the best overall profile. The model has a negative bias towards this university.

\header{Importance}
We expect $imp_{\text{uni}}$ to be higher due to the bias.

\header{Evaluation of the feature attributes}
In Fig.~\ref{figure:synthetic_bar_in_paper}(c), both \method and \greedy are able to identify the negative bias towards one candidate by correctly assigning a higher attribution value to $x_{\text{uni}}$ than for the average query, while \pointwiseShap is not.

\subsection{Highlighting Feature Importance for the Rank of Individual Documents}
\label{section:toy-example-zoom-in}
In this section we zoom in on individual documents and the role of different features on the placement of that documents. For this analysis we use the exposure-based explanation objective from Eq.~\ref{equation:objective_exposure}, highlighting the impact that the different features for the ranking model have on the exposure of the individual candidates. We compare to the attribution values generated by \pointwiseShap for the specific document in question. We investigate two of the scenarios in more detail, the results for the other scenarios can be found in Appendix
B%~\ref{section:appendix_additional_per_query}
.
\footnote{An extended appendix including these additional results is available at \url{https://github.com/MariaHeuss/RankingShap/blob/main/Paper_RankingSHAP.pdf}.} 
Claims made in this subsection on the relative qualities of the candidates can be confirmed with Table~\ref{table:candidate_features} in Appendix~\ref{app:section:experimental-setup}.%\footref{footnote_appendix}

\vspace{-1mm}
\subsubsection{Qualified query}
Since the university and requirements are the same for all candidates, a recruiter might be interested in which features were particularly important for ranking them. \method provides more contrastive insight into the strengths of a document than \pointwiseShap. For example, \method highlights the skill feature as negatively impacting the third candidate's exposure.  If a recruiter is more interested in grades, Fig.~\ref{figure:syn_example_document_based}(a) allows them to make an informed decision to invite the candidate regardless of the model prediction. In contrast, \pointwiseShap provides similar attribution values for each candidate and does not highlight the grades of the third-ranked candidate as a redeeming quality.

\vspace{-1mm}
\subsubsection{Biased query} 
The listwise feature attribution analysis of \method from Fig.~\ref{figure:synthetic_bar_in_paper} shows high importance of the university feature for this query, warranting further investigation. Fig.~\ref{figure:syn_example_document_based}(c) and (d) demonstrate that \method can identify the unfair treatment of the third-ranked candidate due to their university, unlike \pointwiseShap.
\begin{figure}
\centering

\newcommand{\SpacingX}{0.2em}
\newcommand{\Width}{0.22\textwidth}
\newcommand{\Height}{0.21\textwidth}
\newcommand{\BarWidth}{0.0035\textwidth}
\newcommand{\BarWidthI}{0.009\textwidth}
\newcommand{\BarShift}{0.0000\textwidth}

\begin{tikzpicture}
\begin{axis}[
    xbar,
    width=\Width, 
    height=\Height,
    xlabel={},
    symbolic y coords={experience,skills,grades,university,requirements},
    ytick={experience,skills,grades,university,requirements},
    nodes near coords align={horizontal},
    xtick={-0.5,0,0.5}, 
    xmin=-0.55, 
    xmax=0.55,
]
% \addplot [
%     draw=white, % Border color for bars
%     fill=white, % Default fill color for bars
%     bar width=\BarWidth, % Adjust the bar thickness here
% ] table [
%     col sep=comma,
%     x=qualified,
%     y=features,
% ] {figures/csv_files/rankingshap_forth_vs_all_syn.csv};
\addplot [
    draw=mydarkorange, % Border color for bars
    fill=myorange,    
    bar width=\BarWidth, % Adjust the bar thickness here
    bar shift=-1mm, %
] table [
    col sep=comma,
    x=qualified,
    y=features,
] {figures/csv_files/rankingshap_third_vs_all_syn.csv};
\addplot [
    draw=mydarkyellow, % Border color for bars
    fill=myyellow, % Default fill color for bars
    bar width=\BarWidth, % Adjust the bar thickness here
    bar shift=0mm, %
] table [
    col sep=comma,
    x=qualified,
    y=features,
] {figures/csv_files/rankingshap_second_vs_all_syn.csv};
\addplot [
    draw=mydarkred, % Border color for bars
    fill=myred, % Default fill color for bars
    bar width=\BarWidth, % Adjust the bar thickness here
    bar shift=1mm, %
] table [
    col sep=comma,
    x=qualified,
    y=features,
] {figures/csv_files/rankingshap_first_vs_all_syn.csv};
% \addplot [
%     draw=mydarkred, % Border color for bars
%     fill=myred, % Default fill color for bars
%     bar width=\BarWidthI, % Adjust the bar thickness here
% ] table [
%     col sep=comma,
%     x=qualified,
%     y=features,
% ] {figures/csv_files/importance_nan.csv};
\end{axis}
\node[above,font=\large\bfseries, xshift=0] at (current bounding box.north) {Qualified:  \hspace{0.5em} (a) RankingSHAP};
\end{tikzpicture}%
\begin{tikzpicture}
\begin{axis}[
    reverse legend,
    xbar,
    width=\Width, 
    height=\Height,
    xlabel={},
    symbolic y coords={experience,skills,grades,university,requirements},
    ytick={experience,skills,grades,university,requirements},
    yticklabels ={},
    nodes near coords align={horizontal},
    xtick={-0.5,0,0.5}, 
    xmin=-0.7, 
    xmax=0.7,
    legend style={at={(1.5,0.4)},anchor=east, font=\small},  
    legend image code/.code={
        \draw [#1] (0cm,-0.1cm) rectangle (\BarWidthI, \BarWidth); },
]
% \addplot [
%     draw=white, % Border color for bars
%     fill=white, % Default fill color for bars
%     bar width=\BarWidth, % Adjust the bar thickness here
% ] table [
%     col sep=comma,
%     x=qualified,
%     y=features,
% ] {figures/csv_files/pointwise_shap_4_syn.csv};
\addplot [
    draw=mydarkorange, % Border color for bars
    fill=myorange,    
    bar width=\BarWidth, % Adjust the bar thickness here
    bar shift=-1mm, %
] table [
    col sep=comma,
    x=qualified,
    y=features,
] {figures/csv_files/pointwise_shap_3_syn.csv};
\addplot [
    draw=mydarkyellow, % Border color for bars
    fill=myyellow, % Default fill color for bars
    bar width=\BarWidth, % Adjust the bar thickness here
    bar shift=0mm, %
] table [
    col sep=comma,
    x=qualified,
    y=features,
] {figures/csv_files/pointwise_shap_2_syn.csv};
\addplot [
    draw=mydarkred, % Border color for bars
    fill=myred, % Default fill color for bars
    bar width=\BarWidth, % Adjust the bar thickness here
    bar shift=1mm, %
] table [
    col sep=comma,
    x=qualified,
    y=features,
] {figures/csv_files/pointwise_shap_1_syn.csv};
\legend{Cand. 3, Cand. 2, Cand. 1}
\end{axis}
\node[above,font=\large\bfseries, xshift=-6] at (current bounding box.north) {(b) Pointwise SHAP};
% \node[below, color=white] at (current bounding box.south) {Attribution values};
\end{tikzpicture}

\centering

\centering

\renewcommand{\Height}{0.25\textwidth}

\begin{tikzpicture}
\begin{axis}[
    reverse legend,
    xbar,
    width=\Width, height=\Height,
    xlabel={},
    symbolic y coords={experience,skills,grades,university,requirements},
    ytick=data,
    nodes near coords align={horizontal},
    xtick={-0.5,0,0.5}, 
    xmin=-0.55, 
    xmax=0.55,
    ytick={experience, skills, grades, university, requirements},
    % ytick style={draw=none},  
    % yticklabels ={exp,skills,grades,uni,req},
    legend style={at={(1.5,0.4)},anchor=east, font=\small},  
    legend image code/.code={
            \draw [#1] (0cm,-0.1cm) rectangle (\BarWidthI, \BarWidth); },
    ]
]
\addplot [
    draw=mydarkpink, % Border color for bars
    fill=mypink, % Default fill color for bars
    bar width=\BarWidth, % Adjust the bar thickness here
    bar shift=-1.5mm,
] table [
    col sep=comma,
    x=biased,
    y=features,
] {figures/csv_files/rankingshap_forth_vs_all_syn.csv};\label{forth}
\addplot [
    draw=mydarkorange, % Border color for bars
    fill=myorange,    
    bar width=\BarWidth, % Adjust the bar thickness here
    bar shift=-0.5mm,
] table [
    col sep=comma,
    x=biased,
    y=features,
] {figures/csv_files/rankingshap_third_vs_all_syn.csv};\label{third}
\addplot [
    draw=mydarkyellow, % Border color for bars
    fill=myyellow, % Default fill color for bars
    bar width=\BarWidth, % Adjust the bar thickness here
    bar shift=0.5mm,
] table [
    col sep=comma,
    x=biased,
    y=features,
] {figures/csv_files/rankingshap_second_vs_all_syn.csv};\label{second}
\addplot [
    draw=mydarkred, % Border color for bars
    fill=myred, % Default fill color for bars
    bar width=\BarWidth, % Adjust the bar thickness here
    bar shift=1.5mm,
] table [
    col sep=comma,
    x=biased,
    y=features,
] {figures/csv_files/rankingshap_first_vs_all_syn.csv};\label{first}
% \addplot [
%     draw=mydarkred, % Border color for bars
%     fill=myred, % Default fill color for bars
%     bar width=\BarWidthI, % Adjust the bar thickness here
% ] table [
%     col sep=comma,
%     x=biased,
%     y=features,
% ] {figures/csv_files/importance_nan.csv};\label{importance}
\end{axis}
\node[above,font=\large\bfseries, xshift=0] at (current bounding box.north) {Biased: \hspace{1.5em} (c) RankingSHAP};
% \node[below, color=white] at (current bounding box.south) {Attribution values};
\end{tikzpicture}%
\begin{tikzpicture}
\begin{axis}[
    reverse legend,
    xbar,
    width=\Width, height=\Height,
    xlabel={},
    symbolic y coords={experience,skills,grades,university,requirements},
    ytick=data,
    nodes near coords align={horizontal},
    xtick={-0.5,0,0.5}, 
    xmin=-0.7, 
    xmax=0.7,
    yticklabels={},
    % ytick style={draw=none},  
    legend style={at={(1.5,0.4)},anchor=east, font=\small},  
    legend image code/.code={
            \draw [#1] (0cm,-0.1cm) rectangle (\BarWidthI, \BarWidth); },
    ]
]
\addplot [
    draw=mydarkpink, % Border color for bars
    fill=mypink, % Default fill color for bars
    bar width=\BarWidth, % Adjust the bar thickness here
    bar shift=-1.5mm, %
] table [
    col sep=comma,
    x=biased,
    y=features,
] {figures/csv_files/pointwise_shap_4_syn.csv};\label{forth}
\addplot [
    draw=mydarkorange, % Border color for bars
    fill=myorange,    
    bar width=\BarWidth, % Adjust the bar thickness here
    bar shift=-0.5mm,
] table [
    col sep=comma,
    x=biased,
    y=features,
] {figures/csv_files/pointwise_shap_3_syn.csv};\label{third}
\addplot [
    draw=mydarkyellow, % Border color for bars
    fill=myyellow, % Default fill color for bars
    bar width=\BarWidth, % Adjust the bar thickness here
    bar shift=0.5mm,
] table [
    col sep=comma,
    x=biased,
    y=features,
] {figures/csv_files/pointwise_shap_2_syn.csv};\label{second}
\addplot [
    draw=mydarkred, % Border color for bars
    fill=myred, % Default fill color for bars
    bar width=\BarWidth, % Adjust the bar thickness here
    bar shift=1.5mm,
] table [
    col sep=comma,
    x=biased,
    y=features,
] {figures/csv_files/pointwise_shap_1_syn.csv};\label{first}
% \addplot [
%     draw=mydarkred, % Border color for bars
%     fill=myred, % Default fill color for bars
%     bar width=\BarWidthI, % Adjust the bar thickness here
% ] table [
%     col sep=comma,
%     x=biased,
%     y=features,
% ] {figures/csv_files/importance_nan.csv};\label{importance}
\legend{Cand. 4, Cand. 3, Cand. 2, Cand. 1}
\end{axis}
\node[above,font=\large\bfseries, xshift=-6] at (current bounding box.north) {(d) Pointwise SHAP};
% \node[below, color=white] at (current bounding box.south) {Attribution values};
\end{tikzpicture}

\caption{Feature attribution values, for \method with the $g_q^{exp(d)}$ exposure objective defined in Section~\ref{section:method_explanation_objectives} and Pointwise SHAP for individual candidate in the ranked list. \label{figure:syn_example_document_based}}

\end{figure}

\subsection{Discussion}\label{section:experiment-simulated-discussion}

\textbf{The Contrastive Use of Feature Attribution.}
We define the estimated feature importance used in this section's evaluation in a contrastive way, comparing them to other queries as well as to other explanation objectives. 
Prior work~\cite{molnar2023interpreting} suggests that attribution values are hard to interpret in isolation; contextualizing them with other model decisions aids understanding.
The use of different explanation objectives makes feature attribution particularly effective for ranking models: since a model decision involves a complex interplay of various decisions about the relative ordering of documents, contrasting different aspects of the decision allows us to uncover nuances that led to a specific model decision.

\header{Using \method to Identify Biases} 
By comparing attribution values of different queries, we can identify instances where a feature expected to be of moderate importance, such as $x_{\text{uni}}$, impacts the decision more than anticipated. For example, in the biased query, we can detect hints of bias in the explanations in Section~\ref{section:query_scenario_bias}. 
Zooming in on what features are most important for the model to provide the individual candidates with exposure in Section~\ref{section:toy-example-zoom-in}, we see that \method identifies the candidate that got negatively effected by the model bias, as well as qualities that might still speak for them. 

\header{Pointwise vs.\ Listwise Ranking Explanations}
From our synthetic example we see that simply using a pointwise explanation approach to explain listwise ranking decisions fails to consider interactions between the feature values of different documents. 
Features that are important for a high ranking score are assigned a high attribution value, independent of whether they are important for the relative ordering of the list.

\header{Selection is not Attribution} 
While feature selection can be a useful tool for understanding ranking models, more nuanced explanations are sometimes necessary to interpret model decisions. Even if the selection approach correctly identifies the most important features, a feature attribution approach is needed to gain detailed insight into the relative importance of the features impacting for example model bias.

\header{Limitations of White Box Check Evaluation}
We acknowledge the limitations of the qualitative evaluation in this section due to the subjective nature of estimated importance, the synthetic experiment setup, and the limited number of queries investigated. Nevertheless, this section is crucial for providing insights into using listwise feature attribution methods like \method. To complement this qualitative evaluation, we will quantitatively compare \method to a broad range of baselines in Section~\ref{section:experiments}.

\section{Quantitative Feature Attribution Evaluating}\label{section:experiments}

The quantitative evaluation of explanations is a difficult task~\citep{lucic-2021-multistakeholder}. 
In contrast to usual machine learning tasks, where labeled data to benchmark different models can be used for the evaluation, for explanations there is nothing like a \emph{ground truth explanation}. 
Evaluating  feature attribution values in particular is challenging, leading to prior work on evaluating feature attribution often defaults to evaluating the feature selection of the top-$k$ features instead~\cite{rong-2022-consistent}. We will follow this strategy, by defining  Preservation and Deletion Checks~\cite{nauta2023anecdotal} for listwise explanations. 
We pose the following two research questions on the correctness/completeness of the explanations: 
\begin{enumerate*}[label=(\textbf{RQ\arabic*})]
    \item Are explanations
    generated with RankingSHAP faithful to the model decision in terms of overall order of the documents? 
    And
    \item Can \method identify features responsible for the distribution of exposure in the ranked list?
\end{enumerate*}
We describe our experimental setup in  Section~\ref{section:experiments_setup}, our evaluation framework in Section~\ref{section:expriments_evaluation_measures}, and our experimental results in Section~\ref{section:experiments_results}.

\vspace{-1mm}
\subsection{Experimental Setup}
\label{section:experiments_setup}
\vspace{-0.5mm}
\subsubsection{Datasets}
% \header{Datasets}
Following~\cite{singh2021extracting} we consider two datasets from LE\-TOR4.0~\cite{qin2013introducing}. MQ2008 consists of 800 queries with pre-computed query-document feature vectors of dimension 46. The MSLR data set consists of 10k queries with query-document feature vectors of dimension 136. For both, we use the train-val-test split of fold1 and evaluate the explanations on the test data. 

\vspace{-1mm}
\subsubsection{Ranking model}
We use the LightGBM \cite{ke2017lightgbm} to train a listwise ranker with LambdaRank, using NDCG as metric. 

\vspace{-1mm}
\subsubsection{Listwise Explanation Objectives} 
To provide additional evidence for the flexibility of \method we use two different explanation objectives: 
    \textbf{RShapK} uses Kendall's tau objective from Eq.~\ref{equation:objective_kendall} to identify features important for the overall ordering of candidate documents.
    \textbf{RShapW} employs the weighted rank difference objective $g^w$ from Eq.~\ref{equation:objective_weighted} to prioritize documents ranked higher by the model.

\vspace{-1mm}
\subsubsection{Baselines}
We consider the following baselines: 
\begin{description}[leftmargin=*,nosep]
    \item[Random:] Random feature attribution, normalized. 
    
    \item[PWSHAP] Previously used as a baseline in \cite{singh2021extracting}, we take the mean over the pointwise SHAP values of the top-5 documents. 
    
    \item[PWLime:] The mean over the pointwise attribution values generated with LIME of the top-5 documents. 
    
    \item[\greedy:] A greedy feature selection approach from \cite{singh2021extracting}. The authors iteratively add features with the biggest marginal contribution to the initially empty explanation set until a set size of $k$ is reached.
    
    \item[RLime:] Listwise LIME for rankers, inspired by RankLIME~\cite{chowdhury2023rank}. Perturbation is done on each feature of each document independently.
    Since we are interested in listwise explanations, we report the mean of feature attribution values over all documents. 
    
    \item[ShaRP]
    As discussed in Section~\ref{sec:related-work}, parallel to our work, \citet{pliatsika2024sharp} generate feature attribution explanations with SHAP for input features of individual documents, rather than the ranked list as a whole. We use the ``Rank Quantity of Interest'' for our implementation as it is closest in idea to our Kendall-tau based implementation of \method. We use the mean of the individual document explanations to get listwise explanations.
\end{description}

\vspace{-1mm}
\subsubsection{Implementation details}
All approaches, except Random, use background data for masking or perturbing input features. 
For MQ2008, we sample 100 random samples from the training data; for MSLR10k, we sample 20 to compensate for higher feature dimensions. 
For evaluation, we sample a different set of 100 background samples for both datasets. We use the KernelSHAP implementation from the SHAP library~\cite{lundberg2017unified} for \method,  PWShap and ShaRP and the TabularExplainer from the LIME library~\cite{ribeiro2016should} for PWLime and RLime, all with default settings.

\vspace{-1mm}
\subsection{Experimental Evaluation}\label{section:expriments_evaluation_measures}

Due to the lack of ground truth attribution values and evaluation frameworks for rankers, we use the deletion and preservation check strategy~\cite{nauta2023anecdotal} from other machine learning tasks, adapted for ranking. A good explanation should replicate the original model output when non-explained features are masked (Preservation check) and significantly alter the output when important features are removed (Deletion check).

Both checks measure the impact of masking features on the model output, evaluated by a function $v$. We sample masking values $b$ from background data $B$ to substitute for non-explained features, resulting in re-ranked lists $\tilde{\pi}_{e,b}$:
\[
\operatorname{Preservation(e) = \mathbb{E}_{b\sim B} [v(\tilde {\pi}_{e,b})]}.
\]
Similarly, the deletion check applies the mask to the features included in the explanation.

For ranked list outputs, we use Kendall's similarity $\tau$ with the original ranked list $\pi$, hence $v^{ \tau }(\tilde {\pi}_{e,b})= \tau ( \pi, \tilde{\pi}_{e,b})$. These checks align with the validity and completeness criteria in \cite{singh2021extracting}. 
Additionally, we evaluate the alignment of the generated explanations with the original model by measuring the exposure difference between each candidate ranked with the original input and the masked input: $v^{\operatorname{exp-diff}}(\tilde{\pi}_{e,b}) = \sum_{d\in \pi} |\operatorname{exp}(\operatorname{rank}(d|\pi)) - \operatorname{exp}(\operatorname{rank}(d|\tilde{\pi}_{e,b}))|$.
We conduct evaluations at explanation sizes of 1, 3, 5, 7, and 10 and report the mean values over all evaluated queries. 

Note that in this approach, we evaluate feature selection explanations as subsets of features, not attribution values. For feature attribution explanations, we use the top-$k$ features.

\vspace{-1mm}
\subsection{Results}\label{section:experiments_results}
The results with the deletion and preservation checks are presented in Fig.~\ref{fig:deletion_checks_full}.

\input{figures/tex_figures/deletion_check_shrunken}

\header{(RQ1) Are Explanations Generated with RankingSHAP Faithful to the Model Decision in Terms of Overall Order of the Documents} 
To address this research question, we evaluate the \emph{correctness} (how well the explanation aligns with the model's decision) and \emph{completeness} (how much relevant information is captured in the features with the highest attribution values) of the explanations. The preservation check with rank-similarity measures how well the ranked list can be reconstructed using only the most important features identified by each explanation approach. As shown in Fig.~\ref{fig:deletion_checks_full} (a), only the Greedy baseline outperforms \method, which is expected since Greedy is designed to maximize this metric through feature selection explanations. Conversely, the deletion check (b), which involves removing the features with the highest attribution values, reveals that \method outperforms all baselines, including the Greedy and all pointwise baselines. These findings are consistent for the MSLR dataset, as illustrated in Fig.~\ref{fig:deletion_checks_full} (c) and (d). 
Overall using an explanation size of 10 features, we achieve approximately 0.7 rank similarity for the MQ2008 data and 0.6 rank similarity for the MSLR-10k data. In contrast, the rank similarity drops to less than 0.2 and 0.4, respectively, when removing these 10 features with the highest attribution values from the model input. Thus, we answer our first research question in the affirmative: \method is capable of faithfully explaining the model decision.

\header{(RQ2) Can \method Identify Features Responsible for the Distribution of Exposure in the Ranked List} 
We compare explanation approaches using the Preservation Check (Fig.~\ref{fig:deletion_checks_full} (e) and (g)) and the Deletion Check (Fig.~\ref{fig:deletion_checks_full} (f) and (h)), alongside the exposure difference metric $v^\text{exp-diff}$ from Section~\ref{section:expriments_evaluation_measures}. The Preservation Check indicates that the exposure difference decreases for all explanation approaches as the explanation size increases. \method and the Greedy approach perform best in the Preservation Check, reducing the exposure difference by 1/2 to 1/3 compared to the random baseline. In the Deletion Check, \method clearly outperforms all other approaches, producing an exposure difference 3 to 5 times greater, depending on the dataset, when the most important features identified by \method are omitted, as opposed to random features. These findings provide evidence that \method effectively identifies features responsible for the distribution of exposure in a ranked list, thus positively answering the research question.

\vspace{-1mm}
\subsection{Reflections}
\label{section:experiments_discussion}

\header{Using Different Explanation Objectives for Focusing on Different Aspects of the Ranking Decision} 
The performance difference between the two versions of \method, each with distinct explanation objectives, highlights \method's ability to emphasize different aspects of the ranked list for specialized explanations. A listwise similarity objective, like Kendall's tau in RShapK, identifies features critical for the overall ranking. Conversely, an objective like the weighted rank difference in WShapK focuses on the top of the ranked list, improving faithfulness for top documents, as evidenced by exposure-based evaluation. Hence, when using \method for generating ranking explanations, it is crucial to carefully consider which aspects of the ranking decision should be elucidated.

\header{Using SHAP Advances in \method for Enhanced Interpretability}
Since we define \method as a wrapper around SHAP, it is possible to apply improvements developed for SHAP to \method. This allows for the use of numerous advances in the field, such as handling correlated features~\cite{aas2021explaining}, increasing the efficiency of SHAP~\cite{jethani2021fastshap,kariyappa2024shap}, and making adjustments to the sampling of background data~\cite{ghalebikesabi2021locality}, or the weighting of different coalitions when calculating SHAP values~\cite{kwon2022weightedshap}. Some of these advances can be applied directly to \method, although future research will need to investigate how easily transferable these improvements are to the ranking task.
 
\vspace{-1mm}
\section{Conclusion}

In this work, we have defined the concept of listwise feature attribution for ranking tasks, allowing flexible and contrastive examination of ranking decisions through a listwise explanation objective. 
We show that our proposed approach \method results in delivering faithful feature attributions and \method can aid in meaningfully understanding model decisions and detecting biases.

However, we note that \method has limitations, including high computational costs for high-dimensional input spaces and the challenge of interpreting SHAP values, which may not always align with human expectations~\cite{kumar2020problems}, potentially lacking contrastiveness~\cite{miller2019explanation}, and it can be susceptible to adversarial attacks~\cite{slack2020fooling}. 
Additionally, SHAP assumes uncorrelated features, leading to unrealistic out-of-distribution data if ignored~\cite{aas2021explaining}. 
Some of these limitations have been addressed in prior literature, and due to \method's structure as a SHAP wrapper, these improvements could potentially be applied to \method (see Section~\ref{section:experiments_discussion}).

For future work, we see the need for a more thorough evaluation framework that goes beyond faithfulness.
Furthermore, future research should examine whether using listwise SHAP attribution values in a contrastive manner can bridge the gap between mathematically well-defined explanations and practical applications in real-life scenarios.

\header{Data and code}
To facilitate reproducibility, code and parameters are available at \href{https://github.com/MariaHeuss/RankingShap}{https://github.com/MariaHeuss/Ranking\-Shap}.

\vspace{-0.2cm}
\begin{acks}
    We thank Joeri Noorthoek for contributions to the code used in this work, Philipp Hager for feedback on the code, Mathijs Henquet and Jasmin Kareem for feedback on the manuscript.
    This research was (partially) supported by the Dutch Research Council (NWO), under project numbers 024.004.022, NWA.1389.20.\-183, and KICH3.LTP.20.006, the European Union's Horizon Europe program under grant agreement No 101070212,
    the German Research Foundation (DFG), under Project IREM with grant No. AN 996/1-1.
    All content represents the opinion of the authors, which is not necessarily shared or endorsed by their respective employers and/or sponsors.
\end{acks}

\appendix
\label{section:appendix}

\section{Appendix}
\label{app:section:experimental-setup}

Here we include the explicit set-up of the simulated example from Section~\ref{section:experiment-simulated_experiment}. 
In Table~\ref{table:candidate_features} we give an overview over all candidates that were used for the different query scenarios. 
\begin{table}[!h]
    \caption{Feature values for the individual candidates.\label{table:candidate_features}}
    %\footnotesize
    \small
    \centering
    \begin{tabular}{l ccclc}
    \toprule
         % & \text{experience} & \text{skills} & \text{grades} & \text{university} & \text{req}
        candidate & \text{experience} & \text{skills} & \text{grades} & \text{university} & \text{req}
    \\ 
    \midrule
        qual-1 & 0.8 & 0.55 & 3.5 &$\text{uni}_\text{us}$ & True 
    \\
        qual-2 & 0.7 & 0.75 & 3.3 &$\text{uni}_\text{us}$ & True 
    \\
        qual-3 & 0.9 & 0.8 & 3 &$\text{uni}_\text{us}$ & True 
    \\
        non-qual & 0.7 & 0.7 & 3 &$\text{uni}_\text{us}$ & False 
    \\
        privileged & 0.8 & 0.6 & 3.6 & $\text{uni}_\text{nep}$ & False 
    \\
        qual-net & 0.7 & 0.9 & 8 & $\text{uni}_\text{net}$ & True 
    \\
        qual-ger & 0.8 & 0.8 & 1 & $\text{uni}_\text{ger}$ & True
    \\
        qual-biased & 0.8 & 0.7 & 3.6 & $\text{uni}_\text{bias}$ & True
    \\
    \bottomrule
    \end{tabular}
\end{table}
The different universities have different grading schemes, which the models from Figure~\ref{figure:model_flow_chart}
depends on. Table~\ref{tab:university_grades} shows an overview over the different universities that are used in the query scenarios. We show the best possible and the worst passing grade as well as whether the biased model is biased towards the university in question. 
\begin{table}[!h]\small
\centering
\caption{Comparison of grading schemes and model bias across universities.}
\label{tab:university_grades}
\begin{tabular}{lccc}
\toprule
\text{university} & \text{highest grade} & \text{lowest grade} & \text{model bias} \\
\midrule
$\text{uni}_\text{us}$ & 4 & 1 & None \\
$\text{uni}_\text{nep}$ & 4 & 1 & Positive \\
$\text{uni}_\text{bias}$ & 4 & 1 & Negative \\
$\text{uni}_\text{ger}$ & 1 & 4 & None \\
$\text{uni}_\text{net}$ & 10 & 6 & None \\
\bottomrule
\end{tabular}
\end{table} 
Those candidates were then used for different queries. Which candidates were used for what queries can be found in Table~\ref{table:candidate-query-matrix}. 
The table entries indicate the rank of the candidate for the biased ranker, with 0 indicating that they were not included. 
\begin{table}[!h]
    \caption{Query-candidate matrix - numbers indicate the rank for the biased ranker, 0 that they were not considered.\label{table:candidate-query-matrix}}
    \small
    \centering
    \begin{tabular}{l ccccc}
    \toprule
        candidate & average & nepotism & qualified & internat. & biased 
    \\
    \midrule
        qual-1 & 2 & 3 & 3 & 0 & 3 
    \\
        qual-2 & 1 & 2 & 2 & 0 & 2 
    \\
        qual-3 & 0 & 0 & 1 & 2 & 0 
    \\
        non-qual & 3 & 4 & 0 & 4 & 4 
    \\
        privileged & 0 & 1 & 0 & 0 & 0 
    \\
        qual-net & 0 & 0 & 0 & 3 & 0 
    \\
        qual-ger & 0 & 0 & 0 & 1 & 0
    \\
        qual-biased & 0 & 0 & 0 & 0 & 1
    \\
    \bottomrule
    \end{tabular}
\end{table}

\bibliographystyle{ACM-Reference-Format}
\balance
\bibliography{references}
% \clearpage 

% \input{sections/appendix.tex}

\end{document}